\documentclass[twocolumn,tighten,times]{aastex61}

\catcode`\@=11
\newcommand{\gapprox}{\mathrel{\mathpalette\@versim>}}
\newcommand{\lapprox}{\mathrel{\mathpalette\@versim<}}
\newcommand{\propapprox}{\mathrel{\mathpalette\@versim\propto}}
\newcommand{\@versim}[2]
  {\lower3.1truept\vbox{\baselineskip0pt\lineskip0.5truept
\ialign{$\m@th#1\hfil##\hfil$\crcr#2\crcr\sim\crcr}}}
\catcode`\@=12

\received{May 16, 2017}
\accepted{July 28, 2017}
\journalinfo{Astrophysical Journal}

\defcitealias{kaspi17}{KB17}
         
\shorttitle{EXPANSION OF KES 73}

\begin{document}

\title{Expansion of Kes 73, a Shell Supernova Remnant Containing a Magnetar}  

\correspondingauthor{Kazimierz J. Borkowski}
\email{kborkow@ncsu.edu}

\author{Kazimierz J. Borkowski}
\affiliation{Department of Physics, North Carolina State University, 
Raleigh, NC 27695-8202, USA}

\author{Stephen P. Reynolds}
\affiliation{Department of Physics, North Carolina State University, 
Raleigh, NC 27695-8202, USA}

\begin{abstract}

Of the 30 or so Galactic magnetars, about 8 are in supernova remnants
(SNRs).  One of the most extreme magnetars, 1E 1841-045, is at the
center of the SNR Kes 73 (G27.4+0.0), whose age is uncertain.  We
measure its expansion using three {\sl Chandra} observations over 15
yr, obtaining a mean rate of $0.023\% \pm 0.002\%$ yr$^{-1}$.  For a
distance of 8.5 kpc, we obtain a shell velocity of 1100 km s$^{-1}$
and infer a blast-wave speed of 1400 km s$^{-1}$.  For Sedov expansion
into a uniform medium, this gives an age of 1800 yr.  Derived emission
measures imply an ambient density of about 2 cm$^{-3}$ and an upper
limit on the swept-up mass of about 70 $M_\odot$, with lower limits of
tens of $M_\odot$, confirming that Kes 73 is in an advanced
evolutionary stage.  Our spectral analysis shows no evidence for
enhanced abundances as would be expected from a massive progenitor.
Our derived total energy is $1.9 \times 10^{51}$ erg, giving a very
conservative lower limit to the magnetar's initial period of about 3
ms, unless its energy was lost by non-electromagnetic means.  We see
no evidence of a wind-blown bubble as would be produced by a massive
progenitor, or any evidence that the progenitor of Kes 73/1E 1841-045
was anything but a normal red supergiant producing a Type IIP
supernova, though a short-lived stripped-envelope progenitor cannot
be absolutely excluded.  Kes 73's magnetar thus joins SGR 1900+14 as magnetars
resulting from relatively low-mass progenitors.

\end{abstract}


\section{Introduction}
\label{intro}

Magnetars, young neutron stars with surface magnetic fields $B
\gapprox 5 \times 10^{13}$ G, are expected to be born with rotation
periods of order milliseconds and hence with rotational energies
comparable to the energy released in a supernova \citep[see][hereafter
  KB17, for a recent review]{kaspi17}. Their current rotation periods
of 2 -- 12 s are consistent with standard dipole spindown as for
typical rotation-powered pulsars with lower magnetic fields. Much is
uncertain about magnetars, including their progenitors, birth
properties, magnetic-field and thermal evolution, and effects on the
explosions that produce them.  Even the ages of the 30 or so Galactic
magnetars are not well constrained; spindown ages ($P/2{\dot P}$) are
known for about 20, but may not reflect true ages.  If the magnetic
fields result from dynamo action in the neutron-star core, periods of
no more than a few ms are required \citep{duncan92,spruit09}. Fossil
fields have also been proposed as the origin of the high magnetic
fields, e.g., \cite{ferrario06}, but such models have significant
problems accounting for some properties of magnetars \citep{spruit08}. 
Rotation at a few ms periods would be difficult to understand if the
proto-neutron star magnetic field is coupled to an extended envelope,
so for this and other reasons \citep[including the small scale height,
20 -- 30 pc, of their Galactic distribution;][]{olausen14}, 
magnetars are often assumed to come
from massive progenitors that explode as stripped-core supernovae
after heavy mass loss, perhaps distinguishing them from normal pulsars
whose birth periods are thought to be considerably longer
\citep[ca.~100 ms; see, e.g.,][]{gullon15}.

Several magnetars inhabit supernova remnants \citepalias[8 --
  10;][]{kaspi17}, and analysis of the remnants has been brought to
bear to constrain the properties of their magnetars.  \cite{vink06}
showed that for three SNR/magnetar pairs, the deduced explosion
energies were typical for supernovae, constraining initial magnetar
rotation periods to $P_0 \gapprox 5$ ms (unless somehow the magnetar
energy lost is not imparted to the ejecta, for instance as neutrinos
or gravitational radiation).  \cite{martin14} showed that the same
three SNRs do not differ in X-ray luminosity or spectrum (abundances
and ionization state) from remnants hosting normal pulsars.

Since the spindown age is not a reliable indicator of true age for
magnetars \citepalias{kaspi17}, a more direct measure is the
observation of expansion proper motions, possible with {\sl Chandra}
for the expected shock velocities of order 1000 km s$^{-1}$ or
greater.  Here we report observations of the expansion of the X-ray
remnant Kes 73 (G27.4+0.0) which constrain its age.

As documented in \citetalias{kaspi17}, the magnetar in Kes 73, 1E
1841-045, has a period of 11.8 s (the longest of 30 or so), an X-ray
quiescent luminosity of $1.8 \times 10^{35}$ erg s$^{-1}$ between 2
and 10 keV, for a distance of 8.5 kpc \citep{tian08} (second highest,
after SGR 0526$-$66 in the LMC), and a spin-inferred magnetic field of
$7.0 \times 10^{14}$ G, behind only that for SGR 1900+14 and the
recently increased value for SGR 1806-20 \citep[$2 \times 10^{15}$ 
G;][]{younes15}.  Its spindown age of 4600 yr is not atypical. It
has exhibited typical magnetar bursting behavior.  An extended
{\sl Fermi} source includes the location of Kes 73 \citep{li17,yeung17},
but it is four times larger and we consider the association
improbable.

Kes 73 was among the systems examined by \cite{vink06}.  A detailed
X-ray study was performed by \cite{kumar14}, who report an age between
750 and 2100 yr and an explosion energy of about $3 \times 10^{50}$
erg, based on two-component plane-shock spectral fits.  They also
report enhanced abundances of Si and S for one component, which they
attribute to ejecta.

\section{Observations}
\label{obssec}

\begin{deluxetable}{lccc}
\tablecolumns{4}
\tablecaption{{\sl Chandra} Imaging Observations of Kes 73. \label{observationlog}}
\tablehead{
\colhead{Date} & Observation ID & Roll Angle & Effective Exposure Time \\
& & (deg) & (ks) }

\startdata
2000 Jul 23--24   &   729 & 235 & 29.26 \\
2006 Jun 30       &  6732 & 242 & 24.86 \\
2015 Jun 04--05   & 16950 & 129 & 28.65 \\
2015 Jul 07       & 17668 & 201 & 20.88  \\
2015 Jul 08--09   & 17692 & 201 & 23.26 \\
2015 Jul 09--10   & 17693 & 201 & 22.57 \\
\enddata

\end{deluxetable}

{\sl Chandra} observed Kes 73 in 2000 (Epoch I) and 2006 (Epoch II).
Our Epoch III observations took place in 2015
June and July in 4 individual pointings (Table~\ref{observationlog}),
with the remnant again placed on the Advanced CCD Imaging Spectrometer
(ACIS) S3 chip. Observations were reprocessed with CIAO version 4.8
and CALDB version 4.7.1, then screened for periods of high particle
background. Very Faint mode was used;
the total
effective exposure time is 95.4 ks. 

We aligned the first three 2015 pointings to the reference frame of
the 2015 Jul 09--10 observation (Obs.~17693) using photons from the
remnant itself, 
a variation of our more general method for measuring expansion,
described below.
Only an image
extracted from the longest (28.7 ks) pointing from 2015 June
(Obs.~16950) required smoothing; shifts between this and other
pointings were found using unsmoothed images.

The XSPEC spectral analysis package \citep{arnaud96} was used to study
X-ray spectra extracted from individual observations
and added together to obtain merged spectra (response files were
averaged). In our spectral fits, we relied on nonequilibrium
ionization (NEI) models based on APEC NEI version 2.0 atomic data
\citep{smith01}, augmented with inner-shell atomic data as described
in \citet{badenes06} and \citet{borkowski13}.

Archival {\sl Chandra} imaging observations of Kes 73 from 2000 July
and 2006 June (Table~\ref{observationlog}) were reprocessed in the
same way as the Epoch III observations. The Faint mode was employed in these
Epoch I and II
observations instead of the Very Faint mode, so their particle background is
higher than at Epoch III.
As discussed next, their alignment to the 2015 Jul 09--10 observation
was done while measuring expansion.


\section{Expansion} \label{expansion}

\begin{figure}
\hspace*{-3.5mm}
\epsscale{1.3}
\plotone{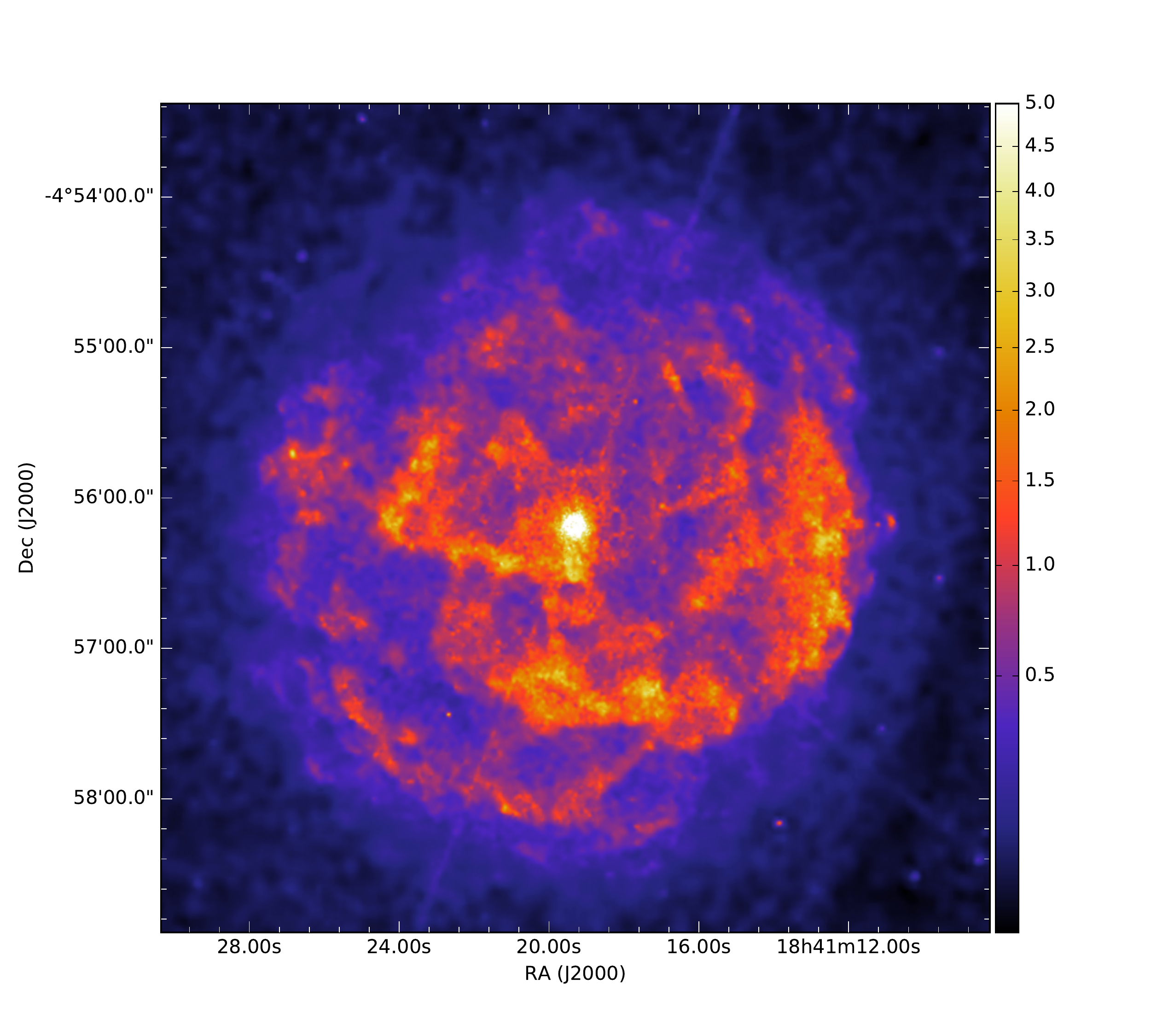}
\caption{Smoothed X-ray image of Kes 73 in the $0.8$--$8$ keV energy range from
2015. The bright magnetar at its center is saturated. 
The scale is in counts per $0 \farcs 323 \times 0 \farcs 323$
image pixel.}
\label{xrays15}
\end{figure}

The time baseline between the 29 ks 2000 and 95 ks 2015 {\sl Chandra}
observations is almost 15 yr.
In order to measure expansion, we use the method of \citet{carlton11}
and \citet{borkowski16}. First, we extracted a data cube from the merged 2015
observations, $1024^2 \times 64$ in size, and in the energy range from 0.8 to
8 keV.
The spatial pixel size is $0 \farcs 323 \times 0 \farcs 323$ (slightly
less than $2/3$ of ACIS $0 \farcs 492$ pixel). We then smoothed this
data cube with the multiscale partitioning method of
\citet{krishnamurthy10}. We set the
penalty parameter that controls smoothing to 0.015, the value that we 
found optimal for G11.2$-$0.3 \citep{borkowski16}.

A smoothed image 
is shown in Figure~\ref{xrays15}. We used this image as a model for 
measuring expansion (after subtraction of background, assumed
to be uniform across the remnant, and normalization by the
monochromatic ($E = 1.49$ keV) exposure map). 
The maximum-likelihood method of \citet{cash79}
was employed to fit the smoothed 2015 image to the earlier raw images
strongly affected by Poisson noise. In addition to varying the
image scale in these fits, we also allowed for changes in the overall
surface brightness by fitting for the surface brightness scale factor
$S$. We also allowed for variations in image shifts in R.A. and
declination, so the image alignment was done simultaneously with
expansion measurements. We accounted for spatial variations in the
effective exposure time at Epochs I and II with
monochromatic exposure maps, and for the higher background rate at
these epochs relative to Epoch III.

\begin{figure}
\hspace*{-4mm}
\epsscale{1.08}
\plotone{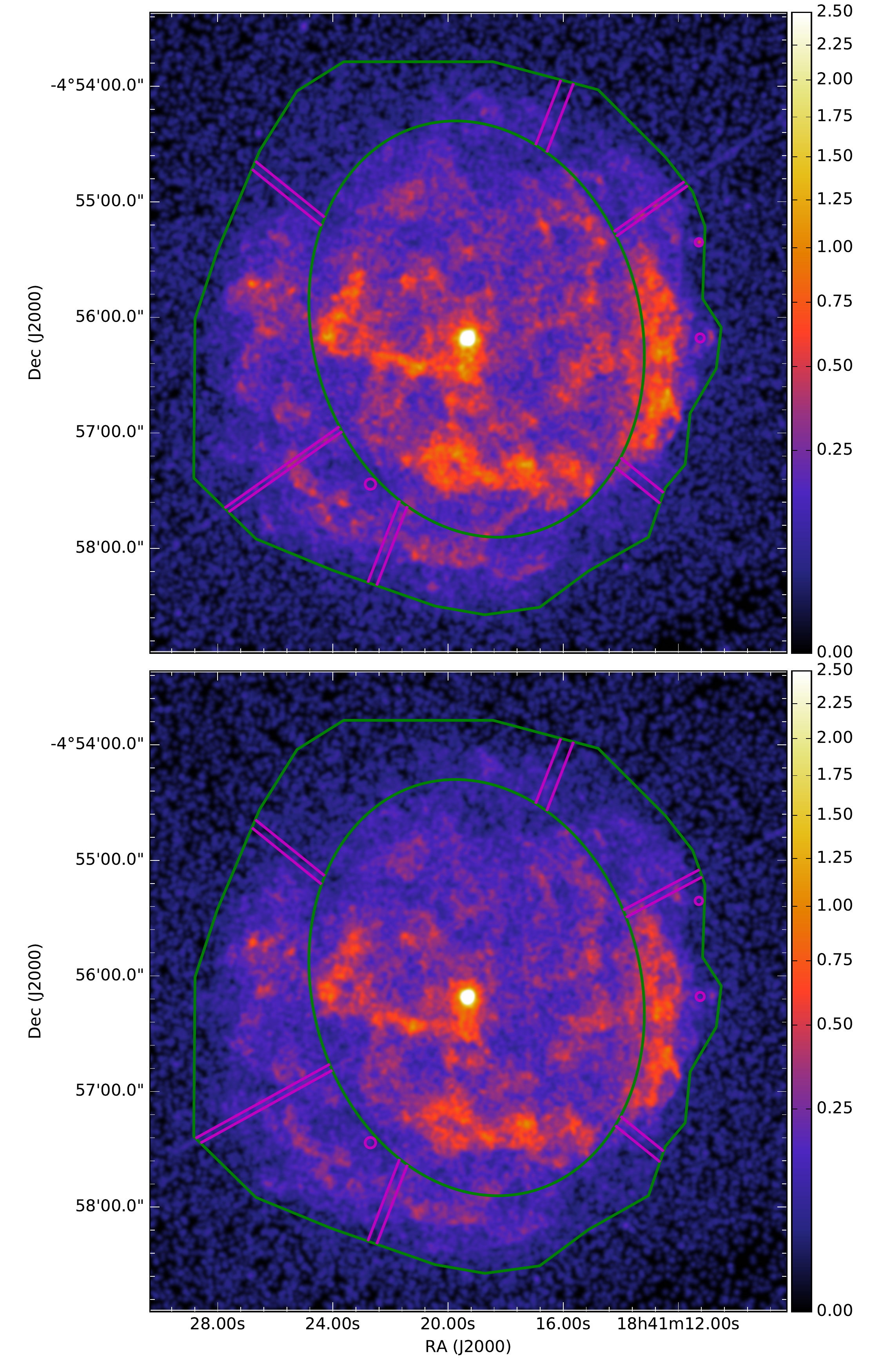}
\caption{ X-ray images of Kes 73 in the $0.8$--$8$ keV energy range
  from 2000 (top) and 2006 (bottom), smoothed with a Gaussian kernel
  with FWHM $=2\farcs3$. The outer shell chosen for expansion measurements is
  demarcated by an ellipse and a polygon (in green), with regions 
  affected by out-of-time events and point sources masked out
  (in magenta). The scales are in counts per
image pixel.}
\label{xrays00and06}
\end{figure}

The spatial morphology of Kes 73 is complex, consisting of overlapping
inner and outer shells, and with bright and clumpy X-ray emission seen
elsewhere within the remnant's interior (Fig.~\ref{xrays15}). The
outer shell most likely traces the location of the blast wave, while
the origin of the interior emission is not clear. Our expansion
measurements are restricted to the outer shell (see masks shown in
Figures \ref{xrays00and06}) where proper motions are
expected to be largest. We assume uniform expansion there, an
assumption that is expected to hold even in the presence of a large
density gradient in the ambient medium \citep{williams13}.
Point sources were excluded. We also excluded regions affected by
``out-of-time'' events caused by photons from the bright, piled-up
magnetar that were detected during the ACIS S3 chip readout, producing
the radial ``streaks'' centered on the magnetar in Figures
\ref{xrays15} and \ref{xrays00and06}.

\begin{deluxetable*}{lccccccc}
\tablecolumns{8}
\tablecaption{Expansion of Kes 73}

\tablehead{
\colhead{Baseline}  & Observation ID & $\Delta t$\tablenotemark{a} & $\Delta \alpha \cos \delta$\tablenotemark{b} & $\Delta \delta$\tablenotemark{c} & $S$\tablenotemark{d} & Expansion  & Expansion Rate \\
\colhead{} & & (yr) &\multicolumn{2}{c}{(arcsec)} & & (\%) & (\%\ yr$^{-1}$) }

\startdata
2000 -- 2015 & 729 & 14.93 & 0.038 & 0.047 & $0.966 \pm 0.005$ & $0.338 \pm 0.033$ & $0.0226 \pm 0.0022$ \\
2006 -- 2015 & 6732 & 8.91 & 0.044 & 0.049 & $0.985 \pm 0.006$ & $0.204 \pm 0.037$ & $0.0229 \pm 0.0042$ \\
(2000+2006) -- 2015 & 729 & 14.93 & 0.038 & 0.046 & $0.966 \pm 0.005$ & \nodata & \nodata \\
& 6732 & 8.91 & 0.043 & 0.048 & $0.985 \pm 0.006$ & \nodata & \nodata \\
& \nodata & \nodata & \nodata & \nodata & \nodata & \nodata & $0.0227 \pm 0.0020$ \\
\enddata

\tablecomments{\ All errors are $1\sigma$.}
\tablenotetext{a}{Baseline length.}
\tablenotetext{b}{Alignment error in R.A.}
\tablenotetext{c}{Alignment error in decl.}
\tablenotetext{d}{Model surface brightness scaling.}
\label{expansiontable}
\end{deluxetable*}

We first measured expansion independently for the Epoch I - III and
Epoch II - III image pairs, with the results listed in the top two
rows of Table~\ref{expansiontable}. Alignment errors are $0 \farcs 04$
in R.A. and $0 \farcs 05$ in decl., so the mean alignment error does
not exceed $0 \farcs 07$, significantly less than the mean
intrinsic error of $0 \farcs 16$ for {\sl Chandra} pointings
\citep{rots09}. The surface brightness scaling factor $S$ is less than
unity for both measurements, 
but only statistical errors are listed in Table~\ref{expansiontable}.
Systematic flux calibration errors are significantly larger than
statistical errors for $S$, most likely at least about $1\%$, so we
find no evidence for temporal variations in either $S$ or the
spatially-integrated flux. Expansion is robustly detected for both
image pairs, at $0.34\% \pm 0.03\%$ for Epochs I and III, and $0.20\%
\pm 0.04\%$ for Epochs II and III. This corresponds to an expansion
rate of $0.023\%$ yr$^{-1}$ in both cases. 

A joint fit to both Epoch I and Epoch II images, with 7 free
parameters (4 image shifts, 2 surface brightness factors $S$, and
expansion rate) instead of 4 for each image pair, gives the most
accurate expansion rate estimate: $0.0227\% \pm 0.0020\%$ yr$^{-1}$,
with other parameters virtually unaffected
(Table~\ref{expansiontable}). Since errors listed in
Table~\ref{expansiontable} include only statistical and not systematic
errors, the total expansion rate error must be somewhat larger than
$0.002\%$ yr$^{-1}$. One source of systematic errors is the smoothing
necessary to obtain the model image shown in
Figure~\ref{xrays15}. This generally biases expansion toward larger
values, and this bias increases with the amount of smoothing. The very
good agreement found in expansion rates for the Epoch I - III and
Epoch II - III image pairs suggests that the magnitude of this bias is
much smaller than statistical errors. This source of systematic errors
was examined by \citet{borkowski16} in their studies of the
significantly brighter SNR G11.2$-$0.3 that expands $20\%$ faster than
Kes 73.  They found it less important than statistical errors, and
this is also expected to hold for Kes 73. 

\section{Blast Wave Speed and Remnant's Age} \label{agespeed}

The mean radius of the outer X-ray shell is $2\arcmin$, so with the
measured expansion rate of $0.023\%$ yr$^{-1}$ we arrive at a shell
velocity of 1100 km s$^{-1}$ at a distance $d$ of 8.5 kpc
\citep{tian08}. \citet{tian08} found lower and upper limits to $d$ to
be 7.5 and 9.8 kpc, respectively, so the shell velocity is between
1000 km s$^{-1}$ and 1300 km s$^{-1}$. There is faint, clumpy X-ray
emission ahead of this shell, so the mean blast wave radius $r_b$ is
larger than $2\arcmin$, and the blast wave speed must be larger than
the mean velocity of the outer shell. Assuming that $r_b$ is equal to
the mean radius of $2 \farcm 5$ for the radio shell \citep{kumar14},
we obtain a blast wave speed $v_b$ of 1400 km s$^{-1}$ at 8.5 kpc,
or 1200 -- 1600 km s$^{-1}$ for $d$ between 7.5 and 9.8 kpc.

The shell expansion rate of $0.023\%$ yr$^{-1}$, assuming no
deceleration, gives an undecelerated remnant's age of 4400 yr. This is
essentially the same age (within errors) as the magnetar's spindown
age of 4600 yr. This undecelerated age is an upper limit to the
remnant's true age $t_{\rm SNR}$, as the blast wave must have been
decelerated: the deceleration parameter $m$ ($r_b \propto t^m$) is
less than unity. A conservative lower limit of 1800 yr is obtained by
assuming Sedov evolution into a constant-density medium (the standard
Sedov model; $m = 2/5$).  However, Kes 73 might still be expanding
into a circumstellar medium (CSM) produced by mass loss from the
progenitor \citep{chevalier05}. Assuming Sedov evolution into a
steady-state wind ($m=2/3$), the lower limit to $t_{\rm SNR}$ becomes
2900 yr.  In principle, the Sedov evolutionary stage, with the ejecta
mass negligible compared to the swept-up mass, might not have been
attained yet; \citet{kumar14} reported the detection of very strong
emission from SN ejecta that, if confirmed, would suggest a
dynamically young remnant (though see below). The deceleration
parameter $m$ could then be larger than the values of $2/5$ or $2/3$
for the two alternatives just mentioned. Based on our expansion
measurements alone, we constrain the age of Kes 73 to be between 2000
and 4000 yr.  (We describe further constraints below.)  A
significantly younger age (even as short as 750 yr) had been
previously thought to be viable
\citep[][]{gotthelf97,vink06,tian08,kumar14}, but only the longest of
these indirect age estimates are consistent with the measured
expansion.

The remnant's age and evolutionary status can be further constrained
with results derived from X-ray spectroscopy.  The blast wave speed of
1400 km s$^{-1}$ we obtain from our direct expansion measurements is
consistent with the values deduced from spectral fits to standard and
wind Sedov models by \citet{kumar14} of $v_b$ of $1200 \pm 300$ km
s$^{-1}$ and $1600 \pm 700$ km s$^{-1}$, respectively, but the much
faster shocks also considered by them are ruled out.  For these
particular models, \citet{kumar14} estimated $t_{\rm SNR}$ at $2100
\pm 500$ yr and $2600 \pm 600$ yr, again consistent with our results.

The swept-up mass $M_{sw}$ in Kes 73 is large:  
\citet{kumar14} estimated $M_{sw}$ at (14 -- 40) $d_{8.5}^{5/2}
M_\odot$ (where $d_{8.5}$ is the distance in units of 8.5 kpc)
assuming a uniform ambient medium, and several times more than this
for the steady-state wind. These mass estimates are based on
two-component spectral fits, with one component assumed to represent
the SN ejecta, and another the shocked ambient medium. Such spectral
decompositions are generally nonunique, and they become even more
unreliable when the ejecta mass $M_{ej} \ll M_{sw}$. Furthermore, previous work
disagrees on the need for elevated abundances; \citet{vink06} examined
the spatially-integrated spectrum of Kes 73 and found no such need for
a two-component model using atomic data from the SPEX package
\citep{kaastra00}, but did require elevated abundances when using an X-ray
Sedov model in XSPEC. The latter was based on the XSPEC NEI version
1.1 atomic data also used by \cite{kumar14}. In order to investigate
this issue further, we extracted a spatially-integrated spectrum from
the 2015 {\sl Chandra} data, and fit it with an X-ray Sedov model with our
improved atomic data. We
found no evidence for enhanced abundances, as setting abundances free
in spectral fits produced solar or even undersolar abundances
depending on the assumed solar abundance set. This resolves the
discrepancy between SPEX- and XSPEC-derived abundances found by
\citet{vink06}. We also extracted spectra from several regions
investigated by \cite{kumar14}, and in disagreement with their results
we again found no evidence for enhanced abundances in fits using more
up-to-date data. Since \citet{kumar14} attributed most of the X-ray
emission to their SN ejecta component, their swept-up mass estimates
(based only on their blast-wave component) must instead be interpreted
as lower limits.

The upper limit to $M_{sw}$ can be estimated from the X-ray emission
measure $EM \equiv \int n_e n_{\rm H} d\,V$ ($n_e$ and $n_{\rm H}$
denote electron and hydrogen number densities, respectively) obtained
from spectral fits assuming solar abundances.  Since there is
presumably some ejecta contribution, even if we do not find obvious
overabundances of heavy elements, the mass we obtain from the total
emission measure gives an upper limit to the swept-up mass.
\citet{vink06} obtained $EM = 2.5 \times 10^{59} d_{8.5}^2$ cm$^{-3}$,
while we find $3.1 \times 10^{59} d_{8.5}^2$ cm$^{-3}$ from fits to
the spatially-integrated {\sl Chandra} spectrum, using an absorbed
Sedov model ({\tt sedov} model in XSPEC), with solar abundances of
\citet{grsa98} assumed for both the X-ray emitting gas and the
(absorbing) ISM.  The good agreement between these independent
estimates demonstrates that $EM$ does not depend much on the spectral
model. Using $EM$ for the standard Sedov model and with $r_b = 2
\farcm 5 = 6.2 d_{8.5}$ pc, we find an upper limit to $M_{sw}$ of $69
d_{8.5}^{5/2} M_\odot$ (this corresponds to the preshock hydrogen
density of $n_0=2.0 d_{8.5}^{-1/2}$ cm$^{-3}$). For the Sedov wind
model, again with $EM = 3.1 \times 10^{59} d_{8.5}^2$ cm$^{-3}$ and
$r_b = 2 \farcm 5$, the upper mass limit becomes $97 d_{8.5}^{5/2}
M_\odot$ \citep[using relationships between $EM$, $r_b$, and $M_{sw}$
  for the Sedov wind model; e.g.,][]{borkowski16}, several times
larger than found by \citet{kumar14} and further strengthening the
case for the advanced dynamical age of Kes 73.  They also strongly
suggest that the blast wave has already entered the ambient ISM;
continuing interaction with CSM, possible only for a very massive
progenitor, is unlikely because of the very large mass of
swept-up CSM required.

We conclude that Kes 73 is in an advanced evolutionary state with
$M_{ej} \ll M_{sw}$ and 
$M_{ej} + M_{sw} \la 70 d_{8.5}^{5/2} M_\odot$, and with the blast
wave now propagating into ambient ISM with a mean speed of
$v_b = 1400 d_{8.5}$ km s$^{-1}$. With an estimated age of
about 2000 yr, it is one of the youngest core-collapse SNRs in our Galaxy.

\section{Discussion}

The mass of the progenitor of Kes 73 and its magnetar is of
considerable interest, as most magnetars are thought to result from
high-mass progenitors.  
The high initial spin frequencies attributed
to magnetars are expected for progenitors that lost most of their
hydrogen envelopes, either through stellar winds or tidal stripping in
close binaries, and finally ended their lives as SNe Ib/c.  Without
extensive envelopes, such stripped-envelope progenitors are less prone
to the spindown experienced by the cores of red supergiant (RSG)
progenitors caused by outward transfer of angular momentum to
overlying hydrogen envelopes \citep{heger05}. Additionally,
progenitors of magnetars in binaries can even be spun up through
mergers, accretion mass transfer, and tidal synchronization \citep[for
a review of scenarios of magnetar formation in binaries,
see][]{popov16}.  Instead of evolving to RSGs, they become compact,
hot, and luminous helium stars burning He in their interiors, although
perhaps with a residual hydrogen envelope still present in some
cases. If sufficiently massive, they are observed as the classical
Wolf-Rayet (WR) stars, while more common, low-mass helium stars in
binary systems are believed to be the progenitors of most SNe Ib.

If Kes 73 resulted from the explosion of a massive star, it should
still be expanding in the progenitor's stellar wind.  The powerful fast
winds of very massive stars, both at the main-sequence (MS) and WR
stage, are very effective in blowing large, low-density bubbles in the
ambient ISM. A star with an initial mass $\ga 30 M_\sun$ may have a
strong enough wind to prevent it from becoming a RSG, particularly if
it is rapidly rotating as expected for magnetar progenitors.  Such a 
star spends virtually all its life blowing fast stellar winds and
inflating its wind-blown bubble to an enormous size. The wind bubble
radius $r_{\rm wb}$ is $0.76 L_w^{1/5} \rho_0^{-1/5} t_w^{3/5}$, where
$L_w$ denotes the wind kinetic luminosity,
$\rho_0 = \mu m_p n_0$ is the ambient
medium density ($\mu=1.4$ denotes the mean mass per hydrogen atom in
atomic mass units, and $m_p$ is the proton mass), and $t_w$ is time
since the onset of the fast wind \citep{weaver77,chevalier89}. For a
rotating $32 M_\sun$ solar-metallicity Geneva stellar model of
\citet{ekstrom12} and \citet{georgy12}, $L_w$ is approximately
constant and equal to $10^{36}$ erg s$^{-1}$ during its entire MS
lifetime of 6.6 Myr \citep{georgy13}. With $n_0 = 2$ cm$^{-3}$,
$r_{\rm wb}$ becomes $70$ pc at the end of the MS stage, and increases
further during its post-MS evolutionary stage that ends in a
long-lived WR star. The small ($6$ pc) radius of Kes 73 and the high
inferred ambient density $n_0 = 2$ cm$^{-3}$ are inconsistent with the
explosion of a very ($\ga 30 M_\sun$) massive star.  Similar arguments
for less massive progenitors, summarized in the Appendix, show
that 20 -- 30 $ M_\sun$ stars also should produce wind bubbles inconsistent
with observations of Kes 73. 

Additional constraints on the SN progenitor can be derived from the lack
of clear evidence for elevated abundances.  The X-ray spectrum of Kes
73 is dominated by strong lines of Mg, Si, and S, elements that are
efficiently produced by massive stars, either through mostly
hydrostatic (Mg) or explosive (Si and S) burning. From our Sedov model
fits, we can estimate how much Mg, Si, and S (mostly swept-up from the
ambient ISM) is present in Kes 73:
$M_{\rm Mg}=0.046 M_\sun$, $M_{\rm  Si}=0.050 M_\sun$, and $M_{\rm S}=0.034 M_\sun$.
For the
nonrotating solar-abundance models of \citet{nomoto13} with an
explosion kinetic energy of $10^{51}$ erg, the Mg yields range from
0.03 -- 0.12 $M_\sun$ for 13 -- 20 $M_\sun$ models, and become much
larger (0.25 -- 0.46 $M_\sun$) for 25 -- 40 $M_\sun$ models.
Similarly, the Si+S yields are 0.10 -- 0.15 $M_\sun$ for the 13 -- 20
$M_\sun$ models, and 0.37 -- 0.54 $M_\sun$ for the 25 -- 40 $M_\sun$
models.  The Si+S yields for stars with lower (13 -- 20 $M_\sun$)
masses only marginally exceed $M_{\rm Si + S} = 0.084 M_\sun$ derived
by us from the Sedov model fit, but they are much larger for stars
with $M > 20 M_\sun$. Only the low ($13 M_\sun$ and $15 M_\sun$) mass
models do not produce more Mg than derived from the observations. If
the ejecta mass is sufficiently large ($\ga 10 M_\sun$, as in the
single-star models just discussed),
there is a possibility that freshly-synthesized Mg, Si, and S might
not have been shocked yet, since it can take a long time for the
reverse shock to arrive at the remnant's center and heat the innermost
ejecta to X-ray emitting temperatures \citep{truelove99}.  But this
requires that the Mg-, Si-, and S-rich ejecta have not been
significantly mixed outwards, an unlikely prospect given that the
magnetar's relativistic wind efficiently removes heavy elements from
the very interior of the SN where they have been produced and mixes
them with the overlying ejecta with the less extreme chemical
composition. The lack of clear chemical enrichment is then more
consistent with an explosion of a relatively low-mass ($\la 20
M_\sun$) progenitor than with a more massive ($\ga 25 M_\sun$) star.

All available evidence points to a relatively low-mass ($< 20 M_\sun$)
progenitor that most likely exploded as a RSG. Long-lived stripped
progenitors appear highly unlikely, including those that originated in low-mass
stars in binaries (see discussion in Appendix). Stripped progenitors are still
possible, but only those that at most underwent only a short-lived
post-RSG phase before exploding as a yellow (or blue) supergiant (or
potentially even as a more compact He star). 

The magnetar's spindown during or shortly after the SN explosion might
have resulted in the injection of a nonnegligible amount of energy
into the SN ejecta, thus contributing to the final supernova kinetic
energy $E$. This energy can now be more reliably estimated assuming
Sedov evolution where $r_b = 12.5 E_{51}^{1/5} n_0^{-1/5} t_4^{2/5}$
pc ($E_{51}$ is $E$ in units of $10^{51}$ erg), with $t_4 \equiv
t_{SNR}/10^4\,{\rm yr}=0.18$ (from our expansion measurements),
$r_b=6.2 d_{8.5}$ pc, and $n_0=2.0 d_{8.5}^{-1/2}$ cm$^{-3}$.
We obtain $E_{51} \approx 1.9 d_{8.5}^{9/2}$, varying from
$1.1$ to $3.5$ with $d$ in the 7.5 -- 9.8 kpc range.

Based on these explosion energy estimates and the lack of conclusive
evidence for ejecta, we conclude that a common (typical) SN IIP could
have produced Kes 73, although the contribution of the magnetar to $E$
is very uncertain as we do not know its initial spin period. The
nature of energy deposition of a magnetar into a SN is not clear at
this time; if it is primarily electromagnetic, almost all of it could
appear as kinetic energy of ejecta \citep{sukhbold17}.  The initial
rotational energy $E_{\rm rot}$ is $2 \times 10^{52} P_{\rm ms}^{-2}$ erg,
where $P_{\rm ms}$ is the initial spin period in ms.  In the extreme
case of the magnetar supplying all of $E$, we find $P_{\rm ms}
\gapprox 3$, a conservative limit.

Our age determination of about 2000 yr rules out the short end of
previously published age ranges for Kes 73. This age is still young
enough that it may pose significant problems for theories of magnetar
cooling or magnetic-field decay \citep[e.g.,][]{vigano13,
beloborodov16}, given the high magnetic field and quiescent X-ray
luminosity of 1E 1841-045.

We conclude that Kes 73 is an undistinguished supernova remnant about
2000 yr old, resulting from an event which is completely consistent
with a run-of-the-mill Type IIP explosion of a red supergiant.  While
\cite{sukhbold17} suggest that magnetars might power ordinary SN IIP
light curves, any RSG phase makes the attainment of ms initial periods
highly problematic \citep{heger05}.  (We note in passing that
\cite{heger05} find initial rotation periods of progenitors of 20
$M_\odot$ or below of 8 -- 18 ms, including neutrino effects: possibly
too slow for magnetars, but too fast to describe the general pulsar
population.) We point out that another magnetar, SGR 1900+14, resulted
from a progenitor of only about 17 $M_\odot$ \citep{davies09}, posing
the same problems as 1E 1841-045.  Binarity may be necessary to remove the
progenitor envelope and allow ms spin periods \citep{turolla15},
though evidence of any surviving companion is scarce.  (\citet{clark14}
report a possible candidate companion for the magnetar CXOU
J164710.2$-$455216.) 1E 1841-045 and SGR 1900+14 may require a
separate formation channel.

\acknowledgments
We acknowledge support by NASA through {\sl Chandra} General Observer
Program grant SAO GO5-16070X. The scientific results reported
here are based on observations made by the {\sl Chandra} X-ray Observatory.
This research has made use of software provided by the
{\sl Chandra} X-ray Center (CXC) in the applications packages {\sl CIAO} and
{\sl ChIPS}. We acknowledge use of various open-source software 
packages for Python, including Numpy, Scipy, Matplotlib, Astropy
(a community-developed core Python package for Astronomy), and
APLpy.\footnote{APLpy is an open-source plotting package for Python hosted at
http://aplpy.github.com.}

\software{CIAO (v 4.8), XSPEC \citep{arnaud96}, SPEX \citep{kaastra00},
  Astropy \citep{astropy13}, matplotlib \citep{hunter07},
  Numpy \citep{walt11}, Scipy \citep{jones01}, APLpy \citep{robitaille12} }

\vspace{5mm}
\facilities{CXO}

\appendix 

Here we describe arguments similar to those in the Discussion above,
to show that somewhat lower-mass progenitors than those discussed
there ($\ga 30 M_\sun$) should also produce wind bubbles
inconsistent with our observations of Kes 73: an ambient density of 2
cm$^{-3}$ and a radius of about 6 pc.  For a $25 M_\sun$ rotating
Geneva model, the brief RSG stage ends long before its explosion
because of extensive mass loss, although its wind-blown bubble might
be much smaller in size than for the more massive stars.
According to \citet{georgy13}, $L_w$ is $\sim$ 3 -- 4 $\times
10^{35}$ erg s$^{-1}$ during its 7.9 Myr-long MS stage. This is likely
an overestimate throughout a significant fraction of its MS evolution
when it can be classified as a dwarf (i.e., luminosity class V) and
its luminosity $L$ is less than $\log L/L_\sun \approx 5.2$, since
below this luminosity threshold observations show much weaker than
expected winds for OB dwarfs \citep[this is known as the weak wind
problem; for a recent review of mass loss in massive stars,
see][]{smith14}. But this problem is not going to affect the final
$15\%$ of the MS stage when its luminosity is above this threshold,
and when at the same time it is no longer expected to be a dwarf
\citep{martins17}. Assuming that the wind starts blowing only after
the first 6.7 Myr, $r_{\rm wb}$ exceeds 20 pc at the end of the MS
stage.
Further growth of the bubble is expected in its post-MS
evolutionary stage, as the star spends $\sim 50\%$ of its time at this stage
blowing a fast stellar wind after evolving to the blue part of the HR diagram
following its brief RSG stage. We conclude that single rotating stars that are
massive enough to spend at most only a small fraction of their post-MS
lifetime as RSGs cannot be progenitors of Kes 73. Since such stars also produce
long-lived classical WR stars, the latter are also excluded.

Rotating Geneva models with $15 M_\sun$ and $20 M_\sun$ spend 1.2 Myr and 0.34
Myr as RSGs, respectively, with the latter's time as an RSG cut short by about 
one half because of extensive mass loss. With typical stellar motions of
5 -- 10 km s$^{-1}$ for young stars, the former can travel up to distances equal
to Kes 73's diameter. This might be enough for the star with $15 M_\sun$
to escape its wind
blown bubble, particularly if its size is small because of a very weak stellar
wind. With $L_w$ about $4 \times 10^{34}$ erg s$^{-1}$ on average during its MS 
lifetime of 13.4 Myr \citep{georgy13}, and accounting for the back pressure
from the ambient ISM that for relatively weak winds significantly decelerates
the bubble's expansion, $r_{\rm wb} = 34$ pc according to equation (4.6) in
\citet{koo92}. Again, this is 
an overestimate because of the weak wind problem that affects such relatively
low-mass stars. A reduction of $L_w$ by 2 orders of
magnitude brings $r_{\rm wb}$ down to
7 pc, but even this scale is
unrealistically high as even a slow (several km s$^{-1}$) stellar motion
will distort the bubble into an elongated, nonspherical tube-like
structure. In
order to investigate this problem further, more realistic bubble models are
needed, with stellar motions included and allowing for a less extreme reduction
in $L_w$, but right now we cannot
exclude the $15 M_\sun$ rotating stellar model as a progenitor for Kes 73 based
on the inferred properties of the swept-up ambient ISM.

The $20 M_\sun$ rotating star is unlikely to be consistent with the inferred
properties of the ambient medium around Kes 73. Because of the weak wind
problem, $L_w$ could possibly be about 2 orders of magnitude less than
the value of $10^{35}$ erg s$^{-1}$ obtained with standard mass loss rates
\citep{georgy13}. But the star is
expected to spend at least $15\%$ of its main sequence lifetime of 9.5 Myr as
a non-dwarf \citep{martins17}. Assuming that the wind with $L_w = 10^{35}$ erg
s$^{-1}$ starts blowing only after the first 8.1 Myr, $r_{\rm wb}$ becomes
18 pc at the end of the MS stage. This is significantly larger than 
Kes 73's radius, although $r_{\rm wb}$ might have been overestimated if the
weak-wind problem persists beyond dwarfs to higher luminosity classes 
\citep{smith17}. After the RSG stage, the star spends
$2 \times 10^5$ yr blowing a fast wind with an average mass-loss rate
$\dot{M}$ of $5 \times  10^{-6} M_\sun$ yr$^{-1}$, eventually becoming a
short-lived WR star. Assuming $v_w = 1000$ km s$^{-1}$, $L_w$ exceeds $10^{36}$
erg s$^{-1}$. Such a strong
wind is capable of blowing a bubble slightly larger than the remnant's size even
if the RSG were able to escape the larger bubble created at the end of the
MS stage. But even if a wind-blown bubble were just a few pc in radius because
of a slightly weaker than assumed wind, a remnant with a particularly prominent
limb-brightened shell is expected after the SN explosion, as observed in 
G11.2$-$0.3 \citep[this CC SNR was inferred to originate from an explosion 
within a small bubble;][]{borkowski16}. The shell of Kes 73 is not as prominent
as in 
G11.2$-$0.3, but further work is needed to investigate whether an explosion in
a small bubble blown shortly before the SN explosion is inconsistent
with the remnant's X-ray and radio morphologies.

Less powerful winds of low-mass He stars in binaries blow smaller
bubbles than the classical WR stars, but their radii are still
substantial.  For a representative low-mass SN Ib progenitor model
with solar abundances, a main-sequence mass of $12 M_\sun$, and a He
core mass $\sim 4 M_\sun$, \citet{gotberg17} estimated $\dot{M}$ at
$2.5 \times 10^{-7} M_\sun$ yr$^{-1}$ during the core He-burning
stage.  All luminous helium stars lose mass in fast ($v_w \sim 1000$
km s$^{-1}$) stellar winds.  Assuming $v_w = 1000$ km s$^{-1}$, $L_w =
8 \times 10^{34}$ erg s$^{-1}$, and again with $n_0 = 2$ cm$^{-3}$,
$r_{\rm wb} = 14 t_{w,{\rm Myr}}^{3/5}$ pc, where $t_{w,{\rm Myr}}$ is
$t_w$ in Myr. The long (1.5 Myr) duration of the He-burning stage
means that a large ($r_{\rm wb} \ga 10$ pc) bubble is expected.  The
small ($6$ pc) radius of Kes 73 and the high inferred ambient density
$n_0 = 2$ cm$^{-3}$ appear inconsistent with an explosion within a
bubble blown by a long-lived low-mass He star progenitor. Still, there
are considerable gaps in our knowledge of these relatively
low-luminosity stars \citep[e.g.,][]{yoon15}, so this conclusion is
not as robust as for the single massive stars that end
their lives as long-lived WR stars.

\end{document}